\documentclass[intlimits,twoside,a4paper]{article}

\usepackage{amsmath,amssymb}
\usepackage{graphicx}

\usepackage[T2A]{fontenc}
\usepackage[cp1251]{inputenc}

\usepackage[eqsecnum]{cmpj2}

\issue{2014}{17}{1}{13001}
\doinumber{10.5488/CMP.17.13001}

\title[Spin-1/2 Ising-Heisenberg diamond chain  with the four-spin interaction]%
{Magnetocaloric effect in the spin-1/2 Ising-Heisenberg diamond chain
 with the four-spin interaction%
}
\author[L. G\'{a}lisov\'{a}]{L. G\'{a}lisov\'{a}}
\address{
Department of Applied Mathematics and Informatics, Faculty of Mechanical Engineering, \\ Technical University in Ko\v{s}ice, Letn\'{a} 9, 042 00 Ko\v{s}ice, Slovak Republic}

\date{Received April 24, 2013, in final form July 12, 2013}
\begin{document}

\maketitle

\begin{abstract}
The magnetocaloric effect in the symmetric spin-1/2 Ising–Heisenberg diamond chain with the Ising four-spin interaction is investigated using the generalized decoration-iteration mapping transformation and the transfer-matrix technique. The entropy and the Gr\"uneisen parameter, which closely relate to the magnetocaloric effect, are exactly calculated to compare the capability of the system to cool in the vicinity of different field-induced ground-state phase transitions during the adiabatic demagnetization.

\pacs 05.50.+q, 75.10.Pq, 75.30.Sg, 75.30.Kz
\end{abstract}

\section{Introduction}
\label{Intro}

The magnetocaloric effect (MCE), which is characterized by an adiabatic change in temperature (or an isothermal change in entropy) arising from the application of the external magnetic field, has been known for more than a hundred years~\cite{War81}. This interesting phenomenon has also got a long history in the cooling applications at various temperature regimes. The first successful experiment of the adiabatic demagnetization, which was used to achieve the temperatures below 1K with the help of paramagnetic salts, was performed in 1933~\cite{Gia33}. Nowadays, the MCE is a standard technique for achieving the extremely low temperatures~\cite{Str07}.

It should be noted that the MCE in quantum spin systems has again attracted much attention of researchers. Indeed, various one- and two-dimensional spin systems have recently been exactly numerically investigated in this context~\cite{Zhi04a,Zhi04b,Hon06,Can06,Der06,Der07,Sch07a,Sch07b,Per09,Can09,Hon09,Tri10,Lan10,Rib10,Jaf12,Top12}.
The main features of the MCE which have been observed during the examination of various spin models include: an enhancement of the MCE owing to the geometric frustration, an enhancement of the MCE in the vicinity of quantum critical points, the appearance of a sequence of cooling and heating stages during the adiabatic demagnetization in spin systems with several magnetically ordered ground states, as well as a possible application of the MCE data for the investigation of critical properties of the system at hand.

In this paper, we investigate the MCE in a symmetric spin-1/2 Ising–Heisenberg diamond chain with the Ising four-spin interaction, which is exactly solvable by combining the generalized decoration-iteration mapping transformation~\cite{Fis59,Syo72,Roj09} and the transfer-matrix technique~\cite{Kra44,Bax82}. As has been shown in our previous investigations~\cite{Gal13}, the considered diamond chain has a rather complex ground state, which predicts the appearance of a sequence of cooling and heating stages in the system during adiabatic demagnetization. The main aim of this work is to compare the adiabatic cooling rate of the system (an enhancement of the MCE) near different field-induced ground-state phase transitions. Bearing in mind this motivation, we investigate the entropy and the Gr\"uneisen parameter during the adiabatic demagnetization process, as well as the isentropes in the $H-T$ plane.

The paper is organized as follows. In section~\ref{sec:model}, we first briefly present the basic steps of an exact analytical treatment of the symmetric spin-1/2 Ising–Heisenberg diamond chain with the Ising four-spin interaction. Exact calculations of the quantities related to the MCE, such as the entropy and the Gr\"uneisen parameter, are also realized in this section. In section~\ref{sec:results}, we briefly recall the ground state of the system, and then the most interesting results for the entropy as a function of the external magnetic field,  the isentropes in the $H-T$ plane and the adiabatic cooling rate of the system versus the applied magnetic field are also presented here. Finally, some concluding remarks are drawn in section~\ref{sec:concl}.

\section{Model and its exact solution \label{sec:model}}

\begin{figure}[ht]
\centerline{\includegraphics[width=7cm]{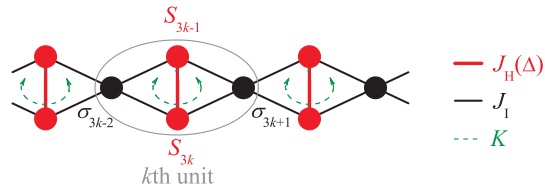}}
\caption{(Color online) A part of the symmetric Ising-Heisenberg diamond chain with the four-spin interaction. The black (red) circles denote lattice positions of the Ising (Heisenberg) spins. The ellipse demarcates spins belonging to the $k$th diamond unit.}
\label{fig1}
\end{figure}

Let us consider a one-dimensional lattice of $N$ inter-connected diamonds in the external magnetic field, which is defined by the Hamiltonian (see figure~\ref{fig1})
\begin{eqnarray}
\label{eq:H}
\hat{{\cal H}}\!\!\!&=&\!\!\! \sum_{k=1}^N\Big[J_{\rm H}\Delta\left(\hat{S}_{3k-1}^{x}\hat{S}_{3k}^{x} + \hat{S}_{3k-1}^{y}\hat{S}_{3k}^{y}\right) + J_{\rm H}\hat{S}_{3k-1}^{z}\hat{S}_{3k}^{z} + J_{\rm I}\left(\hat{S}_{3k-1}^{z}+\hat{S}_{3k}^{z}\right)\left(\hat{\sigma}_{3k-2}^{z}+\hat{\sigma}_{3k+1}^{z}\right) {} \nonumber \\
&&{} + K \hat{S}_{3k-1}^{z}\hat{S}_{3k}^{z}\hat{\sigma}_{3k-2}^{z}\hat{\sigma}_{3k+1}^{z}{}
- H\left(\hat{S}_{3k-1}^{z}+\hat{S}_{3k}^{z}\right)
- H\left(\hat{\sigma}_{3k-2}^{z}+\hat{\sigma}_{3k+1}^{z}\right)/2\Big].
\end{eqnarray}
Here, the spin variables $\hat{S}_{k}^{\gamma}$ ($\gamma = x,y,z$) and $\hat{\sigma}_{k}^{z}$ denote spatial components of the spin-$1/2$ operators, the parameter $J_{\rm H}$ stands for the XXZ Heisenberg interaction between the nearest-neighbouring Heisenberg spins and $\Delta$ is an exchange anisotropy in this interaction. The parameter $J_{\rm I}$ denotes the Ising interaction between the Heisenberg spins and their nearest Ising neighbours, while the parameter $K$ describes the Ising four-spin interaction between both Heisenberg spins and two Ising spins of the diamond-shaped unit. Finally, the last two terms determine the magnetostatic Zeeman's energy of the Ising and Heisenberg spins placed in an external magnetic field $H$ oriented along the $z$-axis.

It is worth mentioning that the considered quantum-classical model is exactly
solvable within the framework of a generalized decoration-iteration mapping transformation~\cite{Fis59,Syo72,Roj09} (for more computational details see our recent works~\cite{Gal13} and~\cite{Can06}). As a result, one obtains a simple relation
between the partition function ${\cal Z}$ of the investigated symmetric spin-1/2 Ising-Heisenberg diamond chain with the four-spin interaction and the partition function ${\cal Z}_{\rm IC}$ of the uniform spin-$1/2$ Ising linear chain with the nearest-neighbour coupling $R$ and the effective magnetic field $H_{\rm IC}$
\begin{eqnarray}
\label{eq:Z}
{\cal Z}(T, J_{\rm I}, J_{\rm H}, K, \Delta, H, N) = A^{N}{\cal Z}_{\rm IC}(T, R, H_{\rm IC}, N).
\end{eqnarray}
The mapping parameters $A$, $R$ and $H_{\rm IC}$ emerging in~(\ref{eq:Z}) can be obtained from the ``self-consistency'' condition of the applied decoration-iteration transformation, and their explicit expressions are given by relations~(4) in reference~\cite{Can06} with the modified $G$ function, which is given by equation~(6) of reference~\cite{Gal13}. It should be mention that the relationship~(\ref{eq:Z}) completes our exact calculation of the partition function because the partition function of the uniform spin-1/2 Ising chain is well known~\cite{Kra44,Bax82}.

At this stage, exact results for other thermodynamic quantities follow straightforwardly. Using the standard relations of thermodynamics and statistical physics, the Helmholtz free energy ${\cal F}$ of the symmetric spin-1/2 Ising-Heisenberg diamond chain with the four-spin interaction can be expressed through the Helmholtz free energy ${\cal F}_{\rm IC}$ of the uniform spin-1/2 Ising chain
\begin{eqnarray}
\label{eq:F}
{\cal F} = -T\ln{\cal Z} = {\cal F}_{\rm IC} - N T\ln A
\end{eqnarray}
(we set the Boltzmann's constant $k_{\rm B}=1$).
Subsequently, the entropy of the investigated diamond chain can be calculated by differentiating the free energy~(\ref{eq:F}) with respect to the temperature $T$. In our case, the resulting equation for the entropy behaves
numerically better if the derivation is taken with respect to the inverse temperature $\beta = 1/T$
\begin{eqnarray}
\label{eq:S}
S = -\!\left(\frac{\partial {\cal F}}{\partial T}\right)_{\!\!H}\!= \beta^2\!\left(\frac{\partial {\cal F}}{\partial \beta}\right)_{\!\!H}\!= \ln{\cal Z_{\rm IC}} + N\ln A - \beta\,\partial_\beta\ln{\cal Z_{\rm IC}} - N\beta\,\partial_\beta\ln A
.
\end{eqnarray}
Here, the functions $\partial_\beta\ln {\cal Z}_{\rm IC}$ and $\partial_\beta\ln A$ satisfy in general the equations
\begin{eqnarray}
\label{eq:partial1}
\partial_x\ln{\cal Z}_{\rm IC} \!\!\!&=&\!\!\!\frac{N}{2}\!\left[\frac{1}{2}+\frac{s^2 - Q^2}{Q(c+Q)} + \frac{s}{Q}\right]\!\partial_x \ln G_- + \frac{N}{2}\!\left[\frac{1}{2}+\frac{s^2 - Q^2}{Q(c+Q)} - \frac{s}{Q}\right]\!\partial_x \ln G_+
\nonumber \\
&&{}-N\!\left[\frac{1}{2}+\frac{s^2 - Q^2}{Q(c+Q)}\right]\!\partial_x \ln G_0 +  \frac{Ns}{2Q}\,\partial_x(\beta H),
\\
\label{eq:partial2}
\partial_x\ln A \!\!\!&=&\!\!\!\frac{1}{4}\left[\,\partial_x \ln G_- + \partial_x \ln G_+ + 2\,\partial_x \ln G_0\,\right]
\end{eqnarray}
with $s=\sinh(\beta H_{\rm IC}/2)$, $c=\cosh(\beta H_{\rm IC}/2)$ and $Q=\sqrt{\sinh^2(\beta H_{\rm IC}/2) + {\rm exp}(-\beta R)}$.
For $x=\beta$ the partial derivatives $\partial_x \ln G_\mp$ and $\partial_x \ln G_0$ emerging in equations~(\ref{eq:partial1}) and~(\ref{eq:partial2}) read
\begin{eqnarray}
\partial_\beta\ln G_\mp \!\!\!&=&\!\!\! \frac{\left(J_{\rm I} \mp H \right)\sinh(\beta J_{\rm I} \mp \beta H)-\left(\frac{J_{\rm H}}{4} + \frac{K}{16}\right)\!\cosh(\beta J_{\rm I} \mp \beta H)}{\cosh(\beta J_{\rm I} \mp \beta H)+\exp\!\left(\frac{\beta J_{\rm H}}{2} + \frac{\beta K}{8}\right)\!\cosh\!\left(\frac{\beta J_{\rm H}\Delta}{2}\right)}
\nonumber\\[1.5ex]
&&{}+
\frac{\left(\frac{J_{\rm H}}{4} + \frac{K}{16}\right)\!\cosh\!\left(\frac{\beta J_{\rm H}\Delta}{2}\right) + \frac{J_{\rm H}\Delta}{2}\sinh\!\left(\frac{\beta J_{\rm H}\Delta}{2}\right)}{{\rm exp}\!\left(-\frac{\beta J_{\rm H}}{2} - \frac{\beta K}{8}\right)\cosh(\beta J_{\rm I} - \beta H)+\cosh\!\left(\frac{\beta J_{\rm H}\Delta}{2}\right)}\,,
\\[1.5ex]
\partial_\beta\ln G_0 \!\!\!&=&\!\!\! \frac{H\sinh(\beta H)-\left(\frac{J_{\rm H}}{4} - \frac{K}{16}\right)\!\cosh(\beta H)}{\cosh(\beta H)+{\rm exp}\!\left(\frac{\beta J_{\rm H}}{2} - \frac{\beta K}{8}\right)\!\cosh\left(\frac{\beta J_{\rm H}\Delta}{2}\right)}
\nonumber\\[1.5ex]
&&{}+
\frac{\left(\frac{J_{\rm H}}{4} - \frac{K}{16}\right)\!\cosh\!\left(\frac{\beta J_{\rm H}\Delta}{2}\right) + \frac{J_{\rm H}\Delta}{2}\sinh\!\left(\frac{\beta J_{\rm H}\Delta}{2}\right)}{{\rm exp}\!\left(-\frac{\beta J_{\rm H}}{2} - \frac{\beta K}{8}\right)\cosh(\beta H)+\!\cosh\!\left(\frac{\beta J_{\rm H}\Delta}{2}\right)}\,.
\end{eqnarray}
Next, let us calculate the quantity called Gr\"uneisen parameter for the investigated model, which closely relates to the MCE. In general, the Gr\"uneisen parameter $\Gamma_H$ can be coupled with the adiabatic cooling rate $(\partial T/\partial H)_{S}$ by using basic thermodynamic relations~\cite{Zhu03,Gar05}:
\begin{eqnarray}
\label{eq:GammaH1}
\Gamma_H = -\frac{(\partial M/\partial T)_H}{C_H} = -\frac{(\partial S/\partial H)_T}{T(\partial S/\partial T)_H} = \frac{1}{T}\!\left(\frac{\partial T}{\partial H}\right)_{\!\!S},
\end{eqnarray}
where $M$ is the total magnetization of the system and $C_H$ is the specific heat at a constant magnetic field $H$. In our case, a direct substitution of the entropy~(\ref{eq:S}) into expression~(\ref{eq:GammaH1}) yields to the following comprehensive form of the Gr\"uneisen parameter $\Gamma_H$ for the symmetric spin-1/2 Ising-Heisenberg diamond chain with a four-spin interaction~(\ref{eq:H}):
\begin{eqnarray}
\label{eq:GammaH2}
\Gamma_H \!\!\!&=&\!\!\!
-\frac{\partial_H\ln{\cal Z}_{\rm IC}+N\partial_H\ln A - \beta\partial^2_{\beta H}\ln{\cal Z}_{\rm IC} - N\beta\partial^2_{\beta H}\ln A}{\beta^2\partial^2_{\beta\beta }\ln{\cal Z}_{\rm IC} + N\beta^2\partial^2_{\beta\beta}\ln A}\,.
\end{eqnarray}
The first two functions $\partial_H\ln{\cal Z}_{\rm IC}$ and $\partial_H\ln A$ occurring in the numerator of the fraction~(\ref{eq:GammaH2}) satisfy the general equations~(\ref{eq:partial1}) and~(\ref{eq:partial2}), respectively, where the derivatives $\partial_x \ln G_\mp$ and $\partial_x \ln G_0$ are given as follows for $x=H$:
\begin{eqnarray}
\partial_H\ln G_\mp \!\!\!&=&\!\!\! \frac{\mp\beta\exp\!\left(-\frac{\beta J_{\rm H}}{2} - \frac{\beta K}{8}\right)\!\sinh(\beta J_{\rm I} \mp \beta H)}{\exp\!\left(-\frac{\beta J_{\rm H}}{2} - \frac{\beta K}{8}\right)\cosh(\beta J_{\rm I} \mp \beta H)+\cosh\!\left(\frac{\beta J_{\rm H}\Delta}{2}\right)}
\,,
\\[1.5ex]
\partial_H\ln G_0 \!\!\!&=&\!\!\! \frac{\beta\exp\!\left(-\frac{\beta J_{\rm H}}{2} - \frac{\beta K}{8}\right)\!\sinh(\beta H)}{\exp\!\left(-\frac{\beta J_{\rm H}}{2} - \frac{\beta K}{8}\right)\cosh(\beta H)+\cosh\!\left(\frac{\beta J_{\rm H}\Delta}{2}\right)}\,.
\end{eqnarray}
Other functions $\partial^2_{\beta H}\ln{\cal Z}_{\rm IC}$, $\partial^2_{\beta H}\ln A$, $\partial^2_{\beta\beta}\ln{\cal Z}_{\rm IC}$ and $\partial^2_{\beta\beta}\ln A$ that emerge in~(\ref{eq:GammaH2}) can be obtained by differentiating~(\ref{eq:partial1}) and~(\ref{eq:partial2}) with respect to $H$ and $\beta$, respectively, provided that $x=\beta$. However, the resulting expressions for these functions are too cumbersome to be written down here explicitly.

\section{Results and discussion}
\label{sec:results}
In this section, we present the results for the entropy as a function of the external magnetic field, isentropes in the $H-T$ plane and the cooling rate during the adiabatic demagnetization for the symmetric spin-1/2 Ising–Heisenberg diamond chain with the Ising four-spin interaction. We assume the Ising and Heisenberg pair interactions $J_{\rm I}$ and $J_{\rm H}$ to be antiferromagnetic ($J_{\rm I}>0, J_{\rm H}>0$), since it can be expected that the magnetic behaviour of the model with the antiferromagnetic interactions in the external longitudinal magnetic field should be more interesting compared to its ferromagnetic counterpart.

\subsection{Ground state}
\label{subsec:GS}
\begin{figure}[htb]
\centerline{\includegraphics[width=1.0\textwidth]{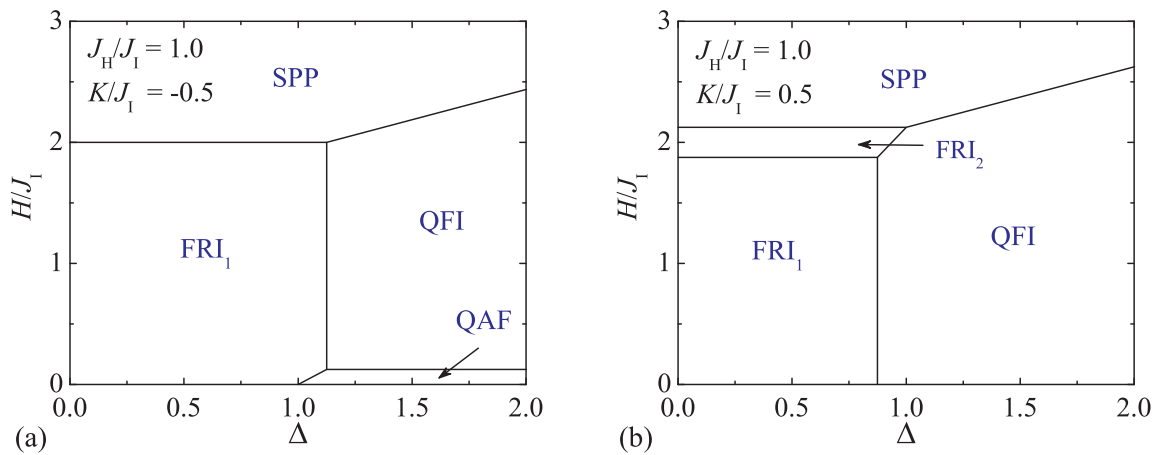}}
\caption{(Color online) Ground-state phase diagrams in the $\Delta - H/J_{\rm I}$ plane for the spin-1/2 Ising–Heisenberg diamond chain with the fixed interaction ratio $J_{\rm H}/J_{\rm I} = 1.0$ and the fixed four-spin interaction (a)~$K/J_{\rm I}=-0.5$, (b)~$K/J_{\rm I}=0.5$.}
\label{fig2}
\end{figure}

In view of a further discussion, it is useful firstly to comment on possible spin arrangements of the investigated diamond chain at zero temperature. Typical ground-state phase diagrams constructed in the $\Delta - H/J_{\rm I}$ plane for the model in the external magnetic field, including all possible ground states, are displayed in figure~\ref{fig2}.
As can be seen from this figure, three different phases appear in the ground state regardless of the nature of the four-spin interaction $K$: the semi-classically ordered ferrimagnetic phase FRI$_1$
 with the perfect antiparallel alignment between the nearest-neighbouring Ising and Heisenberg spins, the quantum ferrimagnetic phase QFI, where all nodal Ising spins occupy the spin state $\sigma^z = 1/2$ and the pairs of Heisenberg spins reside at a quantum superposition of spin states described by the antisymmetric wave function $(|1/2,-1/2\rangle - |-1/2,1/2\rangle)/\!\sqrt{2}$, as well as the saturated paramagnetic phase
SPP, where all Ising and Heisenberg spins are oriented towards the external-field direction. Furthermore, two other interesting phases QAF and FRI$_{2}$ with a perfect antiferromagnetic order in the Ising sublattice can also be found in the ground state depending on whether the four-spin interaction $K$ is considered to be ferromagnetic ($K<0$) or antiferromagnetic ($K>0$), respectively. For more details on the magnetic order of relevant ground states see our recent work~\cite{Gal13}.

\subsection{Entropy}
\label{subsec:S}
\begin{figure}[!h]
\centerline{\includegraphics[width=0.95\textwidth]{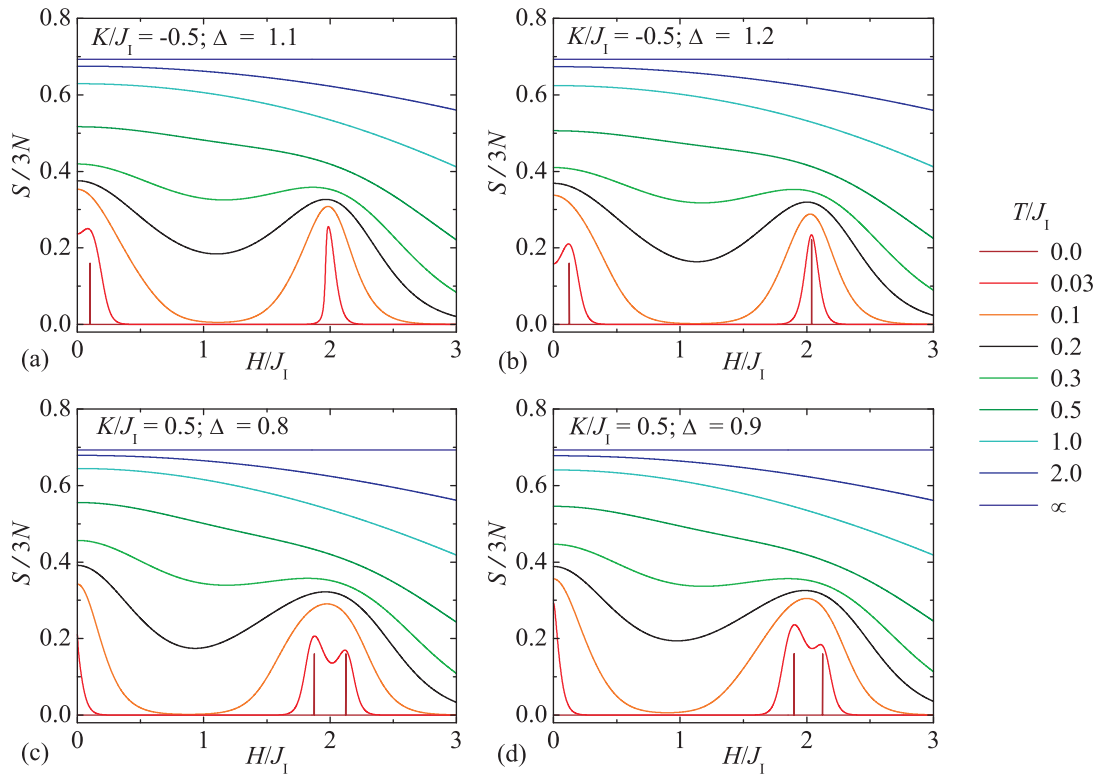}}
\caption{(Color online) Isothermal dependencies of the entropy versus the external magnetic field at various temperatures for the model with the interaction ratio $J_{\rm H}/J_{\rm I} = 1.0$ and the ferromagnetic four-spin interaction $K/J_{\rm I}=-0.5$ when (a)~$\Delta=1.1$, (b)~$\Delta=1.2$, as well as the antiferromagnetic four-spin interaction $K/J_{\rm I}=0.5$ when (c)~$\Delta=0.8$, (d)~$\Delta=0.9$.}
\label{fig3}
\end{figure}

Now, let us turn our attention to the entropy of the investigated diamond chain as a function of the external magnetic field. Figure~\ref{fig3} shows several isothermal dependencies of the entropy per one spin $S/3N$ (recall that the system is composed of $N$ Ising spins and $2N$ Heisenberg spins) versus the magnetic field $H/J_{\rm I}$, corresponding to the spin-1/2 Ising–Heisenberg diamond chain with the fixed interaction ratio $J_{\rm H}/J_{\rm I} = 1.0$ and the fixed ferromagnetic (antiferromagnetic) four-spin interaction $K/J_{\rm I} = -0.5$ ($K/J_{\rm I} = 0.5$). It should be mention that the values of the exchange anisotropy parameter $\Delta$ are chosen so as to reflect all possible field-induced ground-state phase transitions. Evidently, the plotted entropy isotherms are almost unchanged down to temperature $T/J_{\rm I} = 0.5$ for any choice of the parameters $\Delta$ and $K$. In the limit $T/J_{\rm I}\to\infty$, the entropy per spin approaches its maximum value $S_{\mathrm{max}}/3N = \ln 2\approx0.69315$ for any finite value
of the applied magnetic field $H/J_{\rm I}$,
since the spin system is disordered at high temperatures, while it monotonously decreases upon an increase of $H/J_{\rm I}$ when the temperature $T/J_{\rm I}$ is finite. Below $T/J_{\rm I} = 0.5$, the entropy as a function of the magnetic field exhibits irregular dependencies that develop into pronounced peaks located around the transition fields as the temperature is lowered. Finally, almost all these peaks split into isolated lines at critical fields when the temperature reaches the zero value. The only exception is the low-temperature peak observed around the critical field $H_{\mathrm{c}}/J_{\rm I}=2.0$, corresponding to the field-induced phase transition between the phases FRI$_1$ and SPP, which completely vanishes at
$T/J_{\rm I}=0$ [compare the lines for $T/J_{\rm I}=0.03$ and $0$ in figure~\ref{fig3}~(a)]. The residual entropy takes the finite value $S_{\mathrm{res}} = \ln 2$ at this critical point, because just one Ising spin is free to flip in the system and the spin arrangement of its nearest Ising neighbours (and consequently all others) is unambiguously given through the Ising four-spin interaction. Of course, this contribution vanishes in the thermodynamic limit $N\to\infty$, and the residual entropy normalized per spin is $S_{\mathrm{res}}/3N = 0$, which implies that the mixed-spin system is not macroscopically degenerate at the phase transition FRI$_1$--SPP.
However, the macroscopic non-degeneracy of the investigated diamond chain found at $H_{\mathrm{c}}/J_{\rm I}=2.0$ can be observed merely if the four-spin interaction $K$ is ferromagnetic, since the ground-state phase transition FRI$_1$--SPP occurs only for $K<0$ according to the ground-state analysis (see figure~\ref{fig2} as well as figure~2 in the reference~\cite{Gal13}). By contrast, isolated lines appearing in zero-temperature entropy isotherms at other critical fields for $K>0$ as well as $K<0$, whose heights are given by the values of the residual entropy $S_{\mathrm{res}}/3N = \ln 2^{1/3} \approx 0.23105$ and/or $\ln[(1+\sqrt{5})/2]^{1/3} \approx 0.16040$, clearly point to the macroscopic ground-state degeneracy of the system at these points. The former residual entropy $S_{\mathrm{res}}/3N = \ln 2^{1/3}$ found at the ground-state phase transition QFI--SPP is the result of breaking up (forming) the antisymmetric quantum superposition of up-down states of the Heisenberg spins at each unit cell, whereas the latter one $S_{\mathrm{res}}/3N = \ln[(1+\sqrt{5})/2]^{1/3}$ is closely associated with destroying (forming) a perfect antiferromagnetic order in the Ising sublattice at critical fields during the (de)magnetization process.

\subsection{Isentropes and Gr\"uneisen parameter}
\label{subsec:GammaTH}
\begin{figure}[!h]
\centerline{\includegraphics[width=1.0\textwidth]{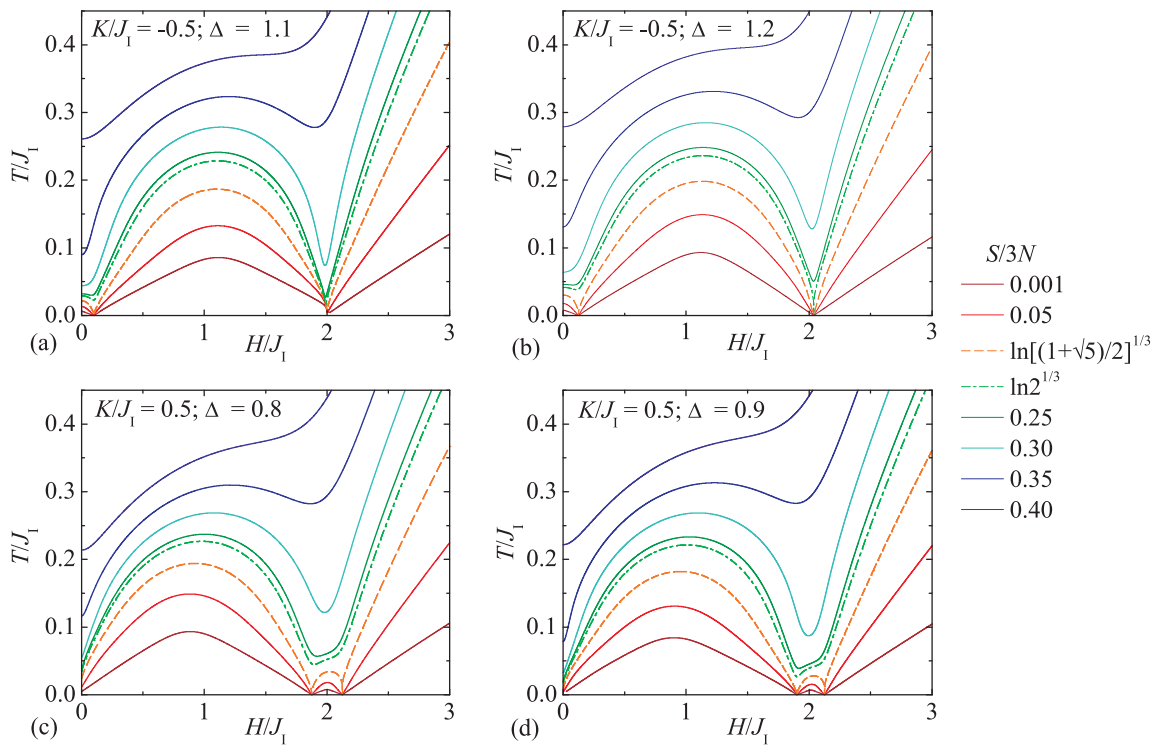}}
\caption{(Color online) The isentropes at the entropy per spin $S/3N=0.001, 0.05,$ \mbox{$\ln[(1+\sqrt{5})/2]^{1/3}, \ln 2^{1/3}, 0.25, 0.3, 0.35$} and $0.4$ in the $H-T$ plane. The values of the interaction parameters  $J_{\rm H}/J_{\rm I}, K/J_{\rm I}$ and $\Delta$ are chosen as in figure~\ref{fig3}.}
\label{fig4}
\end{figure}

In the last part, let us proceed to the investigation of the MCE in its classical
interpretation as an adiabatic change of the temperature of the considered model under field variation. For this purpose, the isentropes in the $H-T$ plane are plotted in figure~\ref{fig4}.
The values of the interaction parameters $J_{\rm H}/J_{\rm I}, K/J_{\rm I}$ and $\Delta$ are chosen as in figure~\ref{fig3}. Comparing figure~\ref{fig4} with ground-state phase diagrams shown in figure~\ref{fig2} one can note that the displayed sets of $T(H)$ curves exhibit a pronounced valley-peak structure, which perfectly reproduces the field-induced phase transitions of the ground state. The most obvious
drop/grow of the temperature can be found in the vicinity of critical fields, where the system undergoes zero-temperature phase transitions. It should be pointed out that this relatively fast cooling/heating of the system near critical points clearly indicates the existence of a large MCE. As can be also found from figure~\ref{fig4}, the temperature of the system reaches the zero value at critical fields if the entropy is less than or equal to its residual value at these points (see also figure~\ref{fig3} showing the isothermal dependencies of the entropy versus the external magnetic field at various temperatures for better clarity).

\begin{figure}[!h]
\centerline{\includegraphics[width=0.9\textwidth]{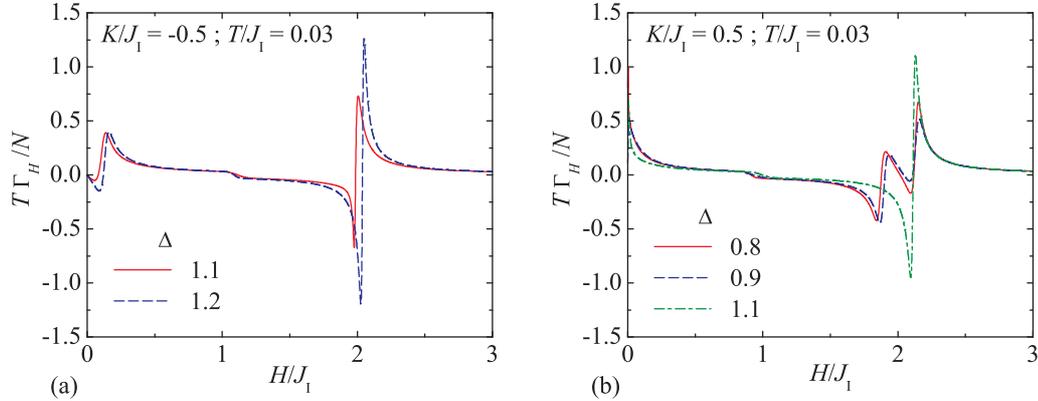}}
\caption{(Color online) The Gr\"uneisen parameter multiplied by the temperature versus the external magnetic field at the fixed temperature $T/J_{\rm I}=0.03$ for the model with the interaction ratio $J_{\rm H}/J_{\rm I}=1.0$ and the set of parameters (a)~$K/J_{\rm I}=-0.5$, $\Delta=1.1, 1.2$; (b)~$K/J_{\rm I}=0.5$, $\Delta=0.8$, $0.9, 1.1$.}
\label{fig5}
\end{figure}
To discuss the adiabatic cooling rate of the system around the ground-state phase transitions in more detail, the Gr\"uneisen parameter multiplied by the temperature $T\Gamma_H$ versus the external magnetic field $H/J_{\rm I}$ at the relatively low temperature $T/J_{\rm I}=0.03$ is depicted in figure~\ref{fig5} by assuming $J_{\rm H}/J_{\rm I}=1.0$, $K/J_{\rm I}=-0.5, 0.5$ and a few values of $\Delta$. Recall that the product $T\Gamma_H$ represents, in fact, the cooling rate $(\partial T/\partial H)_S$ during the adiabatic (de)magnetization [see equation~(\ref{eq:GammaH1})]. As one can see from figure~\ref{fig5}, the displayed low-temperature $T\Gamma_H(H)$ curves pass through the minimum upon increasing the applied magnetic field, change sign from negative to positive values and adopt a maximum within a narrow interval of each critical field, where the system undergoes zero-temperature phase transitions. The sign change in $T\Gamma_H \propto (\partial S/\partial H)_T$ close to the critical fields clearly indicates the presence of a maximum in the corresponding isothermal dependencies of the entropy versus the field in the vicinity of the field-induced ground-state phase transitions (see the lines for $T/J_{\rm I}=0.03$ in figure~\ref{fig3}) and, therefore, we can say that it tracks the accumulation of the entropy due to the competition between neighbouring ground states. Moreover, it is also evident from figure~\ref{fig5} that high-field peaks of the $T\Gamma_H(H)$ curves plotted for the values of $\Delta = 1.2$ in figure~\ref{fig5}~(a) and $\Delta = 1.1$ in figure~\ref{fig5}~(b), emerging at the fields $H/J_{\rm I}\approx2.049$ and $2.129$, respectively, are significantly higher than the others. According to the ground-state phase diagrams shown in figure~\ref{fig2}, these peaks, whose heights are $T\Gamma_H^{\rm peak} \approx 1.26311$ and $1.10248$, appear somewhat above the critical fields associated with the ground-state phase transition QFI--SPP. Other peaks of the heights $T\Gamma_H^{\rm peak} \approx 0.73064$ (see the full red line in figure~\ref{fig5}~(a) for $\Delta = 1.1$), $0.67272$ and $0.51832$ (see full red and dashed blue lines in figure~\ref{fig5}~(b) for $\Delta = 0.8$ and $0.9$, respectively), which can be observed in the field region $H/J_{\rm I}>2.0$, occur just above the phase boundaries FRI$_1$--SPP and FRI$_2$--SPP, respectively. It is thus clear that the cooling effect observed during the adiabatic demagnetization around the ground-state phase transition QFI--SPP is approximately twice of the cooling effect, which can be detected around the phase transitions FRI$_1$--SPP and FRI$_2$--SPP in this $H-T$ range. From these observations one may conclude that the enhancement of the MCE found just around the phase transitions is extremely sensitive to the nature of the degeneracy of the model at these points. Actually, the MCE is the most pronounced around the ground-state boundary QFI--SPP, where strong thermal excitations of the decorated Heisenberg spins are present at low (but non-zero) temperatures due to breaking up the antisymmetric quantum superpositions of their up-down states at $T/J_{\rm I}=0$. By contrast, vigorous low-temperature fluctuations of the Ising spins in the vicinity of other field-induced ground-state phase transitions cause a less pronounced or only a relatively weak cooling effect during the adiabatic demagnetization.

\begin{figure}[!hb]
\centerline{\includegraphics[width=1.0\textwidth]{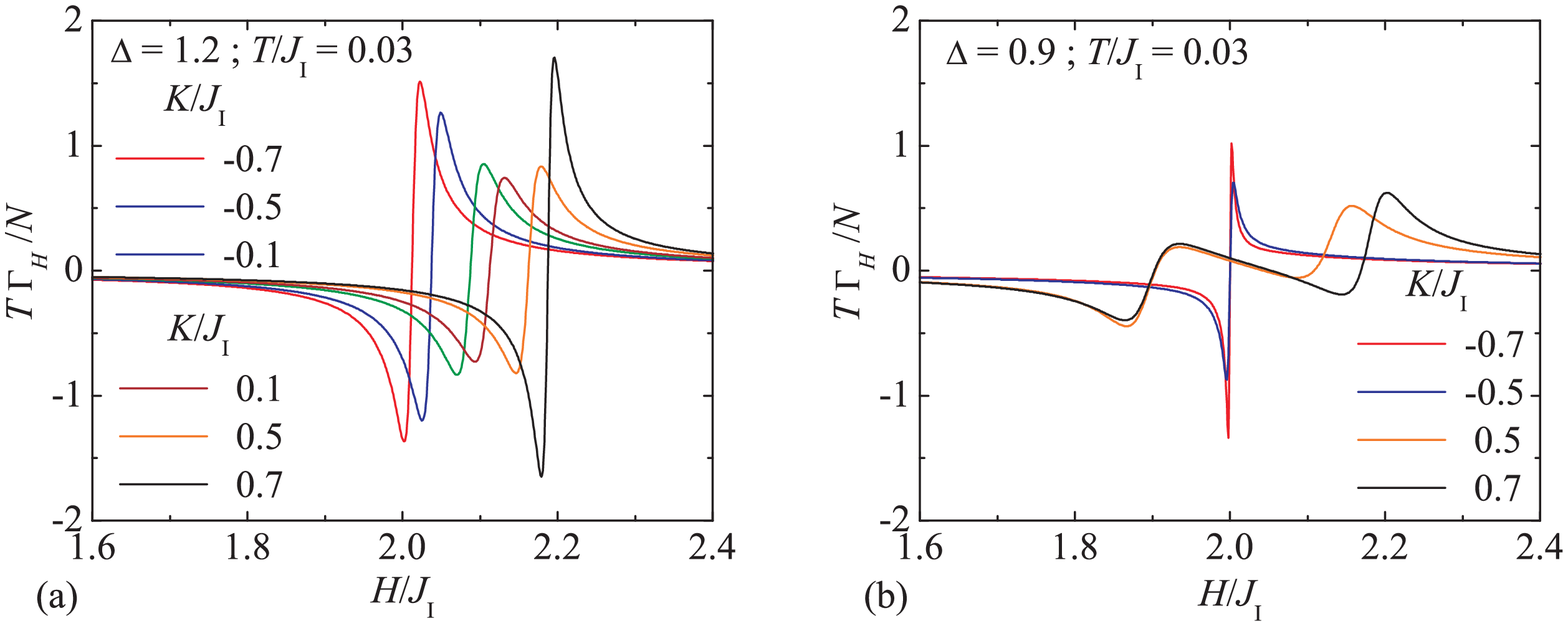}}
\caption{(Color online) The Gr\"uneisen parameter multiplied by the temperature  $T\Gamma_H$ versus the external magnetic field  $H/J_{\rm I}$ at the fixed temperature $T/J_{\rm I}=0.03$ for the model with the interaction ratio $J_{\rm H}/J_{\rm I}=1.0$ and the exchange anisotropy (a)~$\Delta=1.2$; (b)~$\Delta=0.9$, by assuming a few different values of the Ising four-spin interaction $K/J_{\rm I}$.}
\label{fig6}
\end{figure}
\begin{figure}[!h]
\centerline{\includegraphics[width=1.0\textwidth]{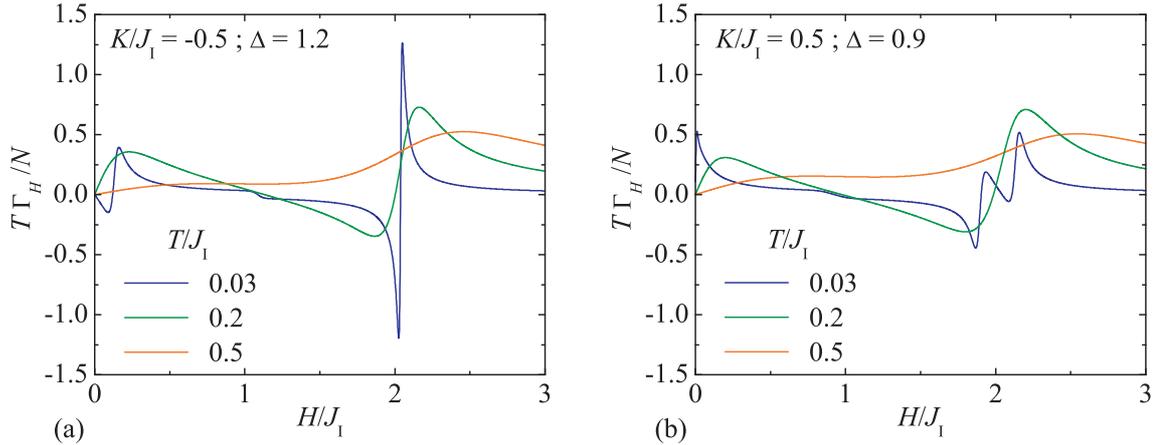}}
\caption{(Color online) The Gr\"uneisen parameter multiplied by the temperature  $T\Gamma_H$ versus the external magnetic field  $H/J_{\rm I}$ for the set of parameters (a)~$K/J_{\rm I}=-0.5$, $\Delta=1.2$; (b)~$K/J_{\rm I}=0.5$, $\Delta=0.9$, by assuming the temperatures $T/J_{\rm I}=0.03, 0.2$ and $0.5$ in both cases.}
\label{fig7}
\end{figure}
The effect of the Ising four-spin interaction on the enhanced MCE in the investigated model is depicted in figure~\ref{fig6}. Figure~\ref{fig6}~(a) demonstrates the situation around the field-induced phase transition QFI--SPP, while figure~\ref{fig6}~(b) shows the situation  around the phase boundaries FRI$_1$--SPP and FRI$_2$--SPP. As can be seen from figure~\ref{fig6}~(a), the peak of the low-temperature $T\Gamma_H(H)$ dependence plotted for $K/J_{\rm I}=-0.7$ is higher than those, which appear in $T\Gamma_H(H)$ curves plotted for $K/J_{\rm I}=-0.5$ and $-0.1$. The similar behaviour can be found for $K>0$: the stronger the antiferromagnetic Ising four-spin interaction $K$ is, the higher peaks can be observed in $T\Gamma_H(H)$ dependencies [see the curves plotted for $K/J_{\rm I}=0.1, 0.5$ and $0.7$ in figure~\ref{fig6}~(a)]. Thus, one may conclude that the adiabatic cooling rate of the system increases with the strengthening of the Ising four-spin interaction $K$ just around the phase boundary QFI--SPP, regardless of its nature. It is clear from figure~\ref{fig6}~(b) that the effect of the interaction $K$ on the adiabatic cooling rate of the system in the vicinity of other field-induced phase transitions FRI$_1$--SPP and FRI$_2$--SPP is the same (note that the peaks of the $T\Gamma_H(H)$ curves plotted for $K/J_{\rm I}=-0.7, -0.5$ appear somewhat above the phase boundary FRI$_1$--SPP, while the peaks of the $T\Gamma_H(H)$ curves plotted for $K/J_{\rm I}=0.5, 0.7$ emerge just above the phase transition FRI$_2$--SPP, see also figure~\ref{fig2}).

For completeness, let us  briefly look at the effect of the temperature on the adiabatic cooling rate of the system. For this purpose, figure~\ref{fig7} illustrates the Gr\"uneisen parameter multiplied by the temperature $T\Gamma_H$ versus the field $H/J_{\rm I}$ for the set of parameters $K/J_{\rm I}=-0.5, \Delta=1.2$ [figure~\ref{fig7}~(a)] and $K/J_{\rm I}=0.5, \Delta=0.9$ [figure~\ref{fig7}~(b)], by assuming three different temperatures.
As expected, the adiabatic cooling rate $T\Gamma_H$ gradually diminishes as the temperature increases. Finally, for sufficiently high temperatures, e.g., for $T/J_{\rm I}=0.5$, the product $T\Gamma_H$ takes only positive values for all magnetic fields, which implies that the thermal fluctuations are already strong enough to drive the system to excited states where no quantum phase transition effects can be seen.

\section{Conclusions}
\label{sec:concl}
In the present paper, we have studied the MCE for the symmetric spin-1/2 Ising–Heisenberg diamond chain with the Ising four-spin interaction, which is exactly solvable by combining the generalized decoration-iteration transformation and the transfer-matrix technique. Within the framework of this approach, we have exactly derived the entropy and Gr\"uneisen parameter, that closely relates to the MCE. We have also obtained the isentropes
in the $H-T$ plane.

We have illustrated that the MCE in the low-entropy and/or low-temperature regimes indicate the field-induced phase transition lines seen in ground-state phase diagrams. More specifically, field-induced ground-state phase transitions perfectly manifest themselves in the form of maxima in low-temperature isothermal dependencies of the entropy versus the external magnetic field, or equivalently in the form of minima in low-entropy isentropes plotted in the $H-T$ plane. This leads to a pronounced cooling of the system during the adiabatic demagnetization in close vicinity of quantum phase transitions when low temperatures are reached. As a consequence, we have found large positive values of the adiabatic cooling rate (the Gr\"uneisen parameter multiplied by the temperature) for magnetic fields slightly above critical points. In addition, we have concluded that the MCE observed just around field-induced ground-state phase transitions is extremely sensitive to the nature of the degeneracy of the model at these points. The most rapid cooling (approximately twice as fast as others) has been observed just around the field-induced ground-state phase transition QFI--SPP, where strong thermal excitations of the decorated Heisenberg spins are present at low temperatures due to breaking up the antisymmetric quantum superpositions of their up-down states at zero temperature, regardless of the nature of the Ising four-spin interaction. By contrast, the effect of Ising four-spin interaction on the adiabatic cooling rate of the system is the same in the vicinity of all field-induced phase transitions. Namely, the increasing Ising four-spin interaction (ferromagnetic as well as antiferromagnetic) accelerates the cooling of the system around phase boundaries during the adiabatic demagnetization.

The considered spin-1/2 Ising–Heisenberg diamond chain with the Ising four-spin interaction, thanks to their simplicity, has enabled the exact analysis of the MCE. Although to our knowledge there is no particular compound which can be described by the  model investigated, our results might be useful in comparing the effects of ground-state phase transitions of different origin on the enhancement of the MCE. On the other hand, the comparison between theory and experiment may be resolved in future in connection with further progress in the synthesis of new magnetic chain compounds.

%
%

\ukrainianpart

\title{Магнетокалоричний ефект у спін-1/2 ромбічноподібному ланцюжку Ізінга-Гайзенберга з чотириспіновою взаємодією}
\author{Л. Г. Ґалісова}
\address{Механіко-інженерний факультет, Технічний університет м. Кошіце,  042 00
Кошіце, Словацька республіка}

\makeukrtitle

\begin{abstract}
\tolerance=3000%
Досліджено магнетокалоричний ефект у симетричному ромбічноподібному ланцюжку Ізінга-Гайзенберга із чотириспіновою взаємодією Ізінга,
використовуючи узагальнене декораційно-ітераційне перетворення і метод трансфер-матриці.
Ентропія і параметр  Грюнайзена, який тісно пов'язаний з магнетокалоричним ефектом, обчислено точно для того, щоб порівняти
здатність системи холонути в околі різних фазових переходів, індукованих полем, під час адіабатичного розмагнічення.
\keywords ромбічноподібний ланцюжок Ізінга-Гайзенберга, чотириспінова взаємодія, фазова діаграма, магнетокалоричний ефект

\end{abstract}

\end{document}